\documentstyle[twoside,fleqn,espcrc2,epsfig]{article}

\newcommand{\no}{\noindent}
\newcommand{\EQ}{\begin{equation}}
\newcommand{\eq}{\end{equation}}
\newcommand{\EQA}{\begin{eqnarray}}
\newcommand{\eqa}{\end{eqnarray}}
\def\pr#1#2#3{ Phys. Rev. {\bf{#1}} (#2) #3}

\def\pl#1#2#3{ Phys. Lett. {\bf{#1}} (#2) #3 }

\def\zp#1#2#3{ Z. Phys. {\bf{#1}} (#2) #3}

\def\epj#1#2#3{ Eur. Phys. J. {\bf{#1}} (#2) #3}

\newcommand{\AmS}{{\protect\the\textfont2
  A\kern-.1667em\lower.5ex\hbox{M}\kern-.125emS}}

\title{Search for Scalar Leptoquarks with polarized
protons (and neutrons) at HERA and future $ep(n)$ Machines}

\author{J.-M. Virey\address{Institut f\"ur Physik, Universit\"at Dortmund,
D-44221 Dortmund, Germany}%
        \thanks{Fellow of the ``Alexander von Humboldt'' Foundation},
        E. Tu\u{g}cu\address{Centre de Physique Th\'eorique, CNRS-Luminy,
Case 907,
F-13288 Marseille Cedex 9, France}\address{Universit\'e de Provence, 
3 Place V. Hugo, F-13331
Marseille cedex 3, France}\address{Galatasaray University, \c 
C\i ra\u gan Cad. 102, Ortak\"oy 80840-\.Istanbul, 
Turkey}
        and
        P. Taxil ${\rm ^b\; }$${\rm ^c}$}

\begin{document}

\begin{abstract}
The effects of Scalar Leptoquarks in various channels have been
analysed for the HERA collider and also for an eventual new
$ep$ machine running at higher energies.
We emphasize the relevance of polarized beams.
\end{abstract}

\maketitle

\section{Introduction}

\vspace{1mm}
\noindent
We present the
effects of Scalar LQ in the Neutral Current (NC) and
Charged Current (CC) channels at HERA, with high integrated luminosities 
and also at an eventual new
$ep$ collider running at higher energies, like the TESLAxHERA
or LEPxLHC projects \cite{Sirois}.
We estimate the constraints that can be reached using those facilities
for several Leptoquark scenarios. We emphasize the relevance
of having polarized lepton and proton beams as well as also having
neutron beams (through polarized $He^3$ nuclei), in order to
disentangle the chiral structure of these various models. 

We adopt the ``model independent'' approach of Buchm\"uller-R\"uckl-Wyler
\cite{BRW} (BRW) where the LQ are classified according to their quantum
numbers and have to fulfill several assumptions like $B$ and $L$
conservation, $SU(3)$x$SU(2)$x$U(1)$
invariance ... (see \cite{BRW} for more details).
The interaction lagrangian is given by :
\EQA
{\cal{L}}&=&\left(g_{1L}\bar{q}_{L}^ci\tau_2\ell_L+g_{1R}\bar{u}_{R}^ce_R
\right). {\bf S_1} + \tilde{g}_{1R}\bar{d}_{R}^ce_R
 . {\bf \tilde{S}_1}\nonumber\\ &+& g_{3L}\bar{q}_{L}^ci
\tau_2\tau\ell_L . {\bf S_3} +
\tilde{h}_2L\bar{d}_R\ell_L . {\bf 
\tilde{R}_2}
\nonumber\\ &+& \left(h_{2L}\bar{u}_R\ell_L+h_{2R}\bar{q}_Li\tau_2
e_R\right) . {\bf R_2} ,
\eqa

\no where the LQ $S_1$, $\tilde{S}_1$ are singlets, $R_2$, $\tilde{R}_2$ are 
doublets and $S_3$ is a triplet. $\ell_L$, $q_L$ ($e_R$, $d_R$, $u_R$)
are the usual lepton and quark doublets (singlets). In what follows
we denote generically by $\lambda$ the LQ coupling and by $M$ the
associated mass. 

These LQ are severely constrained by several different experiments,
and we refer to \cite{LQlim} for some detailled discussions. 

Now, in order to simplify the analysis, we make the following
assumptions : {\it i}) the LQ couple to the first generation only, 
{\it ii}) one LQ
multiplet is present at a time, {\it iii}) the different LQ components
within one LQ multiplet are degenerate in mass, {\it iv}) there is no mixing
among LQ's. From these assumptions and from eq.1, it is possible
to deduce some of the coupling properties of the LQ, which are
summarized in the table 1 of \cite{kalino}.
We stress from this table that the LQ couplings are flavour dependent
and chiral.

\section{Future Constraints}

We consider the HERA collider but with some high integrated
luminosities, namely $L_{e^-}=L_{e^+}=500\, pb^{-1}$. The other
parameters for the analysis being : $e^\pm p$ collisions,
$\sqrt{s}=300\, GeV$, $0.01<y<0.9$, $\left( \Delta\sigma /\sigma
\right)_{syst}\sim 3\,\%$ and GRV pdf set \cite{GRV}. We have considered
also the impact on the constraints of higher energies by considering,
in the one hand, an energy $\sqrt{s}=380\, GeV$ which is closed to the
maximal reach of HERA, and in the other hand, an energy $\sqrt{s}=1\, TeV$ 
which could be obtained at the distant projects TESLAxHERA and/or
LEPxLHC \cite{Sirois}.
Limits at 95\% CL for the various LQ models have been
obtained from a $\chi^2$ analysis performed on the unpolarized NC
cross sections (best observables). 
In figure 1 we compare the sensitivities of various present and future
experiments for $R_{2L}$ as an example. 
\begin{figure}[htb]
\vspace*{-2.2cm}
\centerline{\psfig{figure=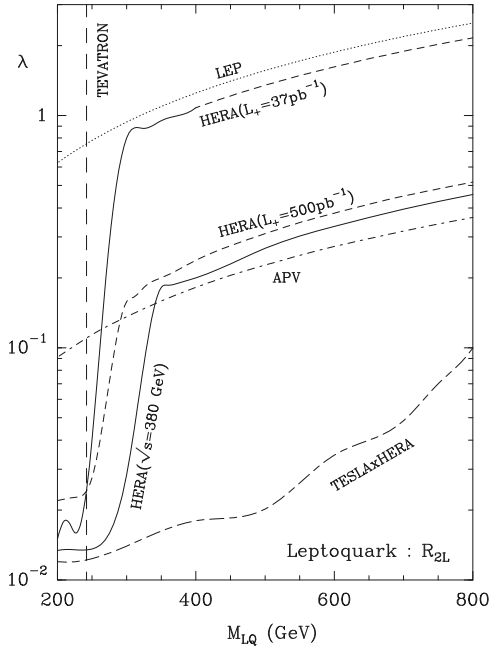,width=8cm}}
\vspace*{-1.5cm}
\caption{Constraints at 95\% CL from various present and future
experiments for $R_{2L}$.}
\label{fig1}
\end{figure}
\vspace*{-0.5cm}
We can remark the followings :
1) LEP limits are already covered by present HERA data \cite{Perez}.
2) For virtual exchange, the LENC constraints (in particular APV 
experiments) are stronger than what could be obtained at HERA even
with higher integrated luminosities and energies.
3) For real exchange, Tevatron data cover an important part of the
parameter space. However, the bounds obtained from LQ 
pair production at Tevatron are strongly sensitive to 
$BR(LQ \rightarrow eq)$ \cite{LQlim}.
It means that there is still an important window
for discovery at HERA in the real domain, especially for more exotic
models like R-parity violating squarks in SUSY models \cite{kalino}.
4) To increase this window of sentivity (for real exchange), it is more
important to increase the energy than the integrated luminosity.
5) A $1\; TeV$ $ep$ collider will give access to a domain (both real and 
virtual) which is unconstrained presently.

\section{Chiral structure analysis}

\subsection{Unpolarized case}

An effect in NC allows the separation of two classes of models. A deviation
for $\sigma^{NC}_{e^-p}$ indicates the class ($S_{1L}$,$S_{1R}$,
$\tilde{S}_{1}$,$S_3$), whereas for $\sigma^{NC}_{e^+p}$ it corresponds
to ($R_{2L}$,$R_{2R}$,$\tilde{R}_{2}$).
For CC events, only $S_{1L}$ and $S_3$ can induce
a deviation from SM expectations (if we do not assume LQ mixing). This
means that the analysis of $\sigma^{CC}_{e^-p}$ can separate the former
class into ($S_{1L}$,$S_3$) and ($S_{1R}$,$\tilde{S}_{1}$). 
If we want to go further into the identification of the LQ
we need to separate "$eu$" from "$ed$" interactions, which seems to be
impossible with $ep$ collisions except if the number of anomalous
events is huge. So, if we want a "complete" separation
of the LQ species we need to consider $ep$ and $en$ collisions as well, 
where some observables like the ratios
of cross sections $R=\sigma^{NC}_{ep}/\sigma^{NC}_{en}$ for instance,
will allow it. However, as soon as we
relax one of our working assumptions ({\it i-iv}) 
some ambiguities will remain. The situation will be better with polarized
collisions.

\subsection{Polarized case}

According to our previous experience \cite{JMV} we know that in general
the Parity Violating (PV) two spin asymmetries exhibit stronger 
sensitivities to new chiral effects
than the single spin asymmetries. Then we consider the case where the
$e$ and $p$ (or neutrons) beams are both polarized.
The PV asymmetries are defined by 
$A_{LL}^{PV} = ({\sigma^{--}_{NC} - \sigma^{++}_{NC}})/(
{\sigma^{--}_{NC} + \sigma^{++}_{NC}})$,
where $\sigma_{NC}^{\lambda_e \lambda_p} \equiv 
(d\sigma_{NC}/dQ^2)^{\lambda_e 
\lambda_p}$, and $\lambda_e, \lambda_p$ are the helicities of the lepton and 
the proton, respectively.
A LQ will induce some effects in these asymmetries, and the directions 
of the deviations from SM
expectations allow the distinction between several classes of models. 
For instance, a positive deviation for $A_{LL}^{PV}(e^-p)$ pins down the 
class ($S_{1L}$,$S_3$)
and, a negative one, the class ($S_{1R}$,$\tilde{S}_{1}$). 
Similarly, an effect for $A_{LL}^{PV}(e^+p)$ makes a distinction between 
the model $R_{2R}$ and the class ($R_{2L}$,$\tilde{R}_{2}$). 
This last fact can be seen in figure 2 which represents $A_{LL}^{PV}$
for $e^+p$ collisions at TESLAxHERA energies with a LQ of mass 500 $GeV$
and coupling $\lambda = 0.2$, the large (small) bars corresponding to 
$L=100(500)\, pb^{-1}$ (a global systematic error of
$\left( \Delta A /A \right)_{syst}=10\,\%$ has been added in quadrature).
\begin{figure}[htb]
\vspace*{-2.2cm}
\centerline{\psfig{figure=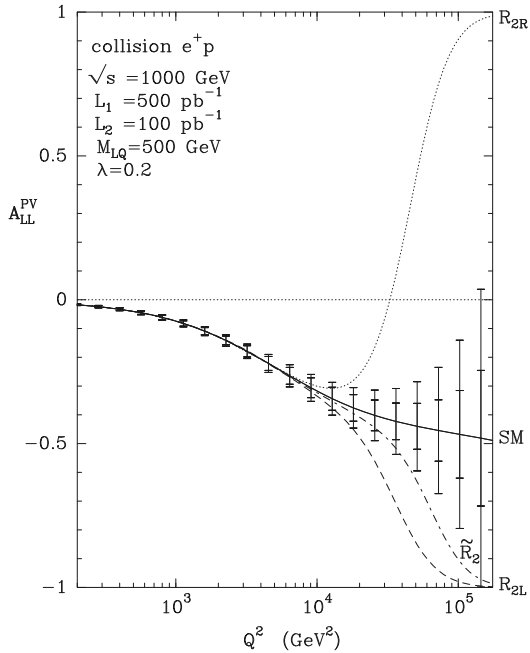,width=8cm}}
\vspace*{-1.5cm}
\caption{$A_{LL}^{PV}(e^+p)$ vs $Q^2$ for the BRW models.}
\label{fig2}
\end{figure}
\vspace*{-0.5cm}

Some other observables, defined in \cite{JMV}, could be used to
go further into the separation of the models.
However the sensitivities of these asymmetries are rather weak, 
and they can
be useful only for some particularly favorable values of the parameters
(M,$\lambda$).
Consequently, polarized $\vec{e}\vec{n}$ collisions are mandatory 
to perform the distinction between the LQ models. This can be seen
through the ratio of asymmetries 
$R ={A_{LL}^{PV}(ep)}/{A_{LL}^{PV}(en)}$,
which for an $e^+$ beam distinguishes the models $R_{2L}$ (positive
deviation) and $\tilde{R}_{2}$ (negative one). This ratio is presented
in figure 3 and the separation
is obvious.
\begin{figure}[htb]
\vspace*{-1.6cm}
\centerline{\psfig{figure=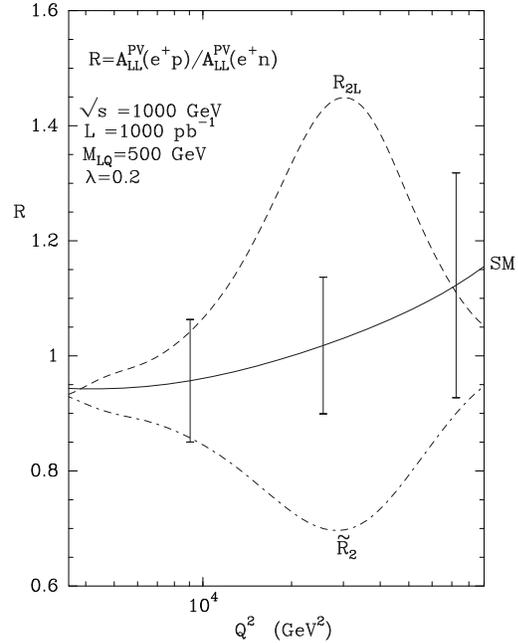,width=8cm}}
\vspace*{-1.5cm}
\caption{$R$ vs $Q^2$ to distinguish $R_{2L}$ and $\tilde{R}_{2}$ models.}
\label{fig3}
\end{figure}
Similarly, for an $e^-$ beam, a positive (negative) deviation in $R(e^-)$ 
indicates the class ($S_{1R}$,$S_{3}$) (($S_{1L}$,$\tilde{S}_{1}$)).
Since these classes are complementary to the ones obtained from
$A_{LL}^{PV}(e^-p)$, it indicates a non-ambiguous separation of the
LQ models. 

%
%
%
%
Finally, if we relax the working assumptions {\it i-iv}, the LQ can have
some more complex structures. Then some ambiguities can remain. Nevertheless, 
the use of additional asymmetries, like the huge number
of charge and PC spin asymmetries that one can define with lepton
plus nucleon polarizations \cite{JMV}, should be very useful 
for the determination of
the chiral structure of the new interaction.

\end{document}